\documentclass[12pt]{article}

\begin{document}                                                     


\def\g{{\tt g}}
\def\1{{\bf 1}}
\def\Z{{\bf Z}}
\def\ee{\end{equation}}
\def\be{\begin{equation}}
\def\l{\label}
\def\D{{\cal D}}
\def\U{{\cal U}}
\def\sin{{\rm sin}}
\def\cos{{\rm cos}}
\def\f{{\bf \Phi}}
\def\v{\varphi}
\def\O{\bf {\cal O}}
\def\C{\bf C}
\def\C{\bf C}
\def\Q{{\cal Q}}
\def\G{{\cal G}}
\def\CP{\bf CP}
\def\e{\rm e}
\def\0{\nonumber}
\def\eea{\end{eqnarray}}
\def\bea{\begin{eqnarray}}
\def\Tr{\rm Tr}
\def\IR{\bf R}
\def\ZZ{\bf Z}
\def\T{\tau}
\def\ep{\epsilon}




\def\title#1{\centerline{\huge{#1}}}
\def\author#1{\centerline{\large{#1}}}
\def\address#1{\centerline{\it #1}}
\def\ack{{\bf Acknowledgments}$\quad$}
\def\Bibliography{\begin{thebibliography}}
\def\endbib{\end{thebibliography}}


\begin{titlepage}

\hfill ULB-TH/04-03

\vspace{1.5cm}

\title{On the boundary gauge dual}
\title{of closed tensionless free strings in AdS}

\vspace{1.5cm}
\author{Giulio Bonelli}
\vspace{.3cm}
\address{
Physique Theorique et Mathematique and International Solvay Institutes}
\address{
 Universite Libre de Bruxelles}
\address{
Campus Plaine C.P. 231; B 1050 Bruxelles, Belgium}
\address{e-mail address: gbonelli@ulb.ac.be}
\begin{abstract}
We consider closed free tensionless strings in $AdS_d$,
calculate exactly the boundary/boundary string evolution kernel
and find the string dynamics to be completely frozen.
We interpret therefore the boundary configurations as Wilson loop
operators in a confining phase. 
This is taken as an argument in favor to the dual weakly coupled 
abelian gauge theory being that of $(d-4)$-forms in the $(d-1)$ dimensional
boundary Minkowski space.
\end{abstract}
\tableofcontents



\end{titlepage}

\section{Introduction}

The tensionless limit of string theory revealed surprising features. 
Its study was started in flat space-time background \cite{G} and was continued in 
\cite{others}, \cite{HT}, \cite{FS} and \cite{iotl}.
In flat space the masslessness of all the higher spin excitation
in the string spectrum actually generates an enhancement of the 
symmetries which reveals \cite{iotl}
in the zero-mode sector being a root concept to understand the tension parameter as 
the most natural length parameter driving a consistent maximal 
symmetry breaking by the mass spectrum lift.
The dynamical description, in the form of a giant ``pantagruelic'' BEH-like phenomenon \cite{massimo}, 
of such a mechanism is of topmost importance to understand the very nature of length scales in string theory.

In AdS/CFT duality, the tensionless limit was conjectured to drive 
the dual gauge Yang-Mills theory to its abelian phase \cite{sw}, but
the first convincing plausibility arguments came with \cite{KP}.
This issues attracted some attention \cite{adscft} and are subject of spread
interest \cite{also}. The connection with higher spin gauge theories
\cite{HS} \cite{HS'} intertwines these two topics.
For more recent results see also \cite{more}.

Since we control much better abelian gauge theories than non abelian ones, this might be 
a better side to study this supposed duality.
The difficult part from the string theory side is not finished, since also string interactions 
are difficult to quantify in curved spaces.
Therefore, it seems that the easiest corner to look for a confirmation of the string/gauge theory duality
is in the tensionless and free limit for closed strings in AdS to be compared with some abelian gauge theory 
in flat Minkowski space. 
The aim of this letter is to focus on that particular narrow corner
and to give some arguments on that conjectured duality.
Actually, the tensionless limit of strings in AdS simplifies very much the theory since
an extra gauge symmetry appears while conformal anomalies are absent due to a particular contraction 
of the Virasoro algebra.
This was developed in \cite{adscov}
while here we apply those results to AdS/CFT duality.

Our approach is quite simple and
follows by considering the AdS space an hyperboloid
in an higher dimensional flat auxiliary space. In such a picture we implement
the restriction to the hyperboloid as a lagrangean constraint to the free theory 
in the ambient space.
The natural set-up for the analysis of such a problem turns out then to be
the constrained hamiltonian formalism \cite{chs} (see also \cite {F} and \cite{HTbook}).
It turns out that the constraint algebra structure simplifies in the 
massless/tensionless case
in such a way that the arising of a larger gauge symmetry takes place.
More specifically, it happens that the geometric constraints appear as second class 
for generic values of the tension parameter, but in the tensionless limit 
it is possible to single out one half of them which, 
together with the tensionless Virasoro generators, are first class.
This property opens the way to a BRST covariant quantization of the system
which was given in \cite{adscov}.

In few words, the main point is that massless free excitations on AdS can not probe
the strength of the space-time curvature and therefore, in this case, the assignment
of the value of a finite AdS radius, being a non physical datum, should be regarded as a gauge choice
(the assignment of a zero radius being a degenerate gauge fixing condition).
Let us underline again that this property is special of massless free excitations 
on AdS space-time.

For other approaches to the problem, see \cite{CKKYLZ} and \cite{ST}.
In particular, in \cite{ST} the covariantization of the string field equations
to negative curvature spaces is obtained.

In this letter we describe explicitly the bulk/boundary correspondence for
tensionless free closed strings. The AdS bulk theory is that of closed tensionless free strings
which is argued to correspond to an abelian Yang-Mills gauge theory
on the boundary Minkowski space. This is obtained by calculating the boundary-boundary
evolution path-integral for a closed string and showing that it reproduces 
the correlation function of two charge conjugate Wilson loop operators 
for an abelian Yang-Mills gauge theory at strong coupling. 
By this we mean just that the correlation vanishes unless the two loops coincide
showing those as carrying a magnetic monopole charge frozen in the confining phase and having 
therefore no dynamics. 
Its electromagnetic dual
weakly coupled abelian theory 
(seen as the zero coupling limit of a non-abelian one)
is than that of a (d-4)-form which gets
indicated by our result, in agreement with \cite{KP},
as the weakly coupled gauge dual of the tensionless free closed string in $AdS_{d}$.
Since the technique is quite model independent,
our argument can be naturally extended to supersymmetric strings.

Let us notice that our resulting picture partly resides on 
an earlier conjecture by Polyakov, pushed forward 
in \cite{Polyakov97} and in \cite{PolyakovS}, about gauge/string theory duality.

In the next section we partially fix the gauge of the
tensionless bosonic closed free strings in AdS in a way which is 
compatible with the boundary conditions on $\partial AdS$ in the string path integral.
In section three we describe two pictures of the resulting boundary theory
and give a plausible weakly coupled abelian gauge theory dual.
A last section is left for few comments.

\section{A partial gauge fixing}

Let us model the dimension $d$ AdS space as an hyperboloid in a flat $d+1$ 
dimensional space. 
Labeling the coordinates in ${\bf R}^{d+1}$ as $x^\mu$, as $\mu=0,\dots,d$, 
the embedding equation is simply 
$$
x^\mu\eta_{\mu\nu}x^\nu=
-x_0^2+x_1^2+\dots+x_{d-1}^2-x_d^2=-R^2
$$
which defines the symmetric quadratic form $\eta$. We will usually write
$UV=U^\mu\eta_{\mu\nu}V^\nu$ for various (co)vectors.

Let us start reviewing the constrained system of closed strings in ${\bf R}^{d+1}$
bound to stay on the hyperboloid $x^2=-R^2$ in the tensionless limit. 

The starting point is the $\alpha'$ expansion of the string coordinates and momenta.
This is designed in the ambient space as
\be
X^\mu(\sigma)=x^\mu+\sqrt{\alpha'}A^\mu(\sigma)
\quad {\rm and}\quad
P_\mu(\sigma)=\frac{1}{2\pi}p_\mu +\frac{1}{\sqrt{\alpha'}}B_\mu(\sigma)
\l{exp}\ee
where $(x,p)$ are the zero modes and $(A,B)$ are the oscillator parts
which are periodic in $[0,2\pi]$, are dimensionless and independent on 
$\alpha'$.

The study of this system amounts to the extraction of the leading order 
terms in $\alpha'$ out of the Virasoro constrains
\be
\frac{1}{2}\left[2\pi\alpha'P^2+\frac{1}{2\pi\alpha'}(X')^2\right]=0
\quad {\rm and}\quad 
X'P=0
\l{virasoro}\ee
and the geometric constrains
\be
\frac{1}{\alpha'}\left(X^2+R^2\right)=0
\quad {\rm and}\quad 
XP=0.
\l{geometric}\ee
In \cite{adscov} it was shown that one can extract one half of the geometric
constrains such that, after the tensionless limit has been taken, together with the contracted 
Virasoro this full set is a 
choice of a complete set of first class constraints for the system.
Their implementation in a quantum BRST charge to give a well defined quantum
theory proves that the above procedure is well defined quantum mechanically.

The most interesting consequence of what we said above is that the left half of the geometric constraints
in the tensionless limit plays the role of a gauge fixing condition rather than having a 
physical content. Because of the result obtained in \cite{adscov}, we can work safely
with the constrained functional path-integral methods where potential
anomalies (from normal ordering ambiguities) could have been be less transparent at a first stage.

In our expansion (\ref{exp}), the contracted Virasoro are equivalent to
\be
p^2=0\quad {\rm and}\quad \oint A'B=0
\l{cvir1}\ee
as far as the zero modes are concerned, while for the oscillators we have
\be
pA=0 \quad {\rm and}\quad pB=0 .
\l{cvir2}\ee
The chosen half of the geometric constraints (namely, the zero mode of $XP$
and the fluctuation of $X^2+R^2$)
are then
\be
xp+\oint AB=0
\quad{\rm and}\quad AA'=0
\l{cgeom}\ee

In our parameterization, to reach the AdS boundary, one has to choose a null
hyperplane 
\footnote{We use adapted light-cone coordinates so that as usual
$UV=U_+V_-+U_-V_++uv$.} 
at will, lets say $X_++\rho=0$,
intersect it with the hyperboloid and then let the hyperplane position $\rho$
go to infinity.

Since our aim is to calculate the boundary-boundary string propagator, we can view 
the hyperplane placement as a partial gauge fixing, compatible with the boundary conditions, 
for our gauge theory.
More completely, we partially gauge fix to
\footnote{At a first stage the reader might be worried by the impression that we are 
forgetting some important geometrical data about the AdS space.
Actually we are just making a finite gauge transformation to the system
which is nothing else than a change in the explicit gauge fixing conditions.}
\be
X_++\rho=0
\quad {\rm and}\quad
P_++\eta=0
\l{pgf}\ee
where $\rho$ and $\eta$ are constants in $\sigma$.
The above conditions (\ref{pgf}) can be rephrased using (\ref{exp}) just as
$x_++\rho=0$, $A_+=0$, $p_++\eta=0$ and $B_+=0$.

As it is clear from a simple degrees of freedom counting, the conditions (\ref{pgf})
do not fix the gauge symmetry completely.
The important aspect of the above partial gauge choice is that it 
fixes completely the degrees of freedom in the two directions
which are transverse to the boundary of the AdS space and let us with a string path integral
in the relevant $d-1$ Minkowski space.

In order to see this in details, we consider the following subset of the gauge algebra generators
\be
p^2 \quad {\rm and}\quad xp+\oint AB
\l{pgfd1}\ee
out of the zero modes and
\be
pA \quad {\rm and}\quad pB
\l{pgfd2}\ee
for the oscillators.
Notice that we have chosen to fix all the contracted Virasoro generators but the level matching 
condition
(which generates constant translations in $\sigma$ and is too important to get rid of)
and the bulk global dilatation generator.

The generators (\ref{pgfd1}) and (\ref{pgfd2}) get fixed completely
by the gauge fixing conditions (\ref{pgf}).
Using the standard Faddeev formula for constrained path-integrals
\cite{F} (see also \cite{HTbook}),
this can be verified by calculating the determinant of the matrix of Poisson
brackets 
\footnote{We refer to the standard Poisson bracket 
$\left\{X^\mu(\sigma),P_\nu(\sigma')\right\}_{PB}=\delta^\mu_\nu\delta(\sigma,\sigma')$,
etc. expanded as in (\ref{exp}).}
to be non zero for generic values of $(\eta,\rho)$
and to get canceled completely by the factors needed to solve 
the delta-functions for the gauge fixed constraints in $(x_-,p_-)$
and $(A_-,B_-)$.
Moreover, it turns out that the matrix of Poisson brackets of the left over gauge
generators, i.e. $\oint BA'$ and $AA'$, with the partial gauge fixing conditions
is zero once the gauge fixing conditions themselves are fulfilled.
The reduced action for the gauge fixed directions, 
corresponding to the chosen boundary conditions in the path integral, 
is $\int_{t_i}^{t_f} d\tau \oint \left(P^+\dot X_+-X^-\dot P_+\right)$
and vanishes once  the gauge fixing conditions (\ref{pgf}) are taken into account.
Notice that the resulting path integral is completely independent on the 
actual values of $(\eta,\rho)$.
Therefore, after this gauge fixing operation is completed, we are left with
the string path-integral along the $d-1$ left coordinates and momenta
with the first class constraints
\be
\oint ba'=0 \quad {\rm and} \quad aa'=0
\l{bgg}\ee
where we denoted with $(a,b)$ the $d-1$ dimensional components of $(A,B)$
others than $(A_\pm,B_\pm)$. The scalar product is taken with the transverse
Minkowskian metric with signature $(1,d-2)$.

\section{Two (conjugate) boundary pictures}

After the previous section, we can now recollect the left over string
theory path integral from the boundary point of view.

The Hamiltonian is made out of gauge generators only, and reads
$$H=\oint \left[\frac{\lambda}{2\pi} ba'+\Lambda aa'\right]$$
where $\lambda$ is a constant Lagrange multiplier while, conversely, $\oint \Lambda=0$.
There are actually two relevant pictures, corresponding to two different types of 
boundary conditions, to discuss.
In them both
the center of mass variables are completely blocked at zero Hamiltonian
and therefore have no evolution\footnote{This is in fact the fate that a 
massless free particle would have.
For the massless particle the relevant path integral is
$$
\int D[X,P]{\rm exp}\left\{i\int d\tau\left[ \dot X_+P_- -\dot P_+X_- +\dot x p\right]\right\}
\delta(P^2)\delta(XP)\delta(P_++\eta)\delta(X_++\rho)
[det(*)]
$$
where $[det(*)]$ is the determinant of the matrix of 
Poisson brackets of the gauge generators and the gauge fixings
which gives $2\eta^2$. This factor is exactly the one needed to solve the $\delta$-functions
in the light cone variables. The result is then independent on the value of $(\rho,\eta)$
and gives just $\delta(x_i-x_f)$, where $x_i$ and $x_f$ are the initial and final transverse positions.
The string case is a generalization of this approach.
A careful reader might like to have a more rigorous proof of the gauge equivalence
of the light-cone model to the one in which the hyperboloid constraint is manifest.
This follows by considering the family of gauge fixing conditions 
$(X^2+R^2)(1-t)+(P_++\eta)t=0$ with $t\in[0,1]$ and $X_++\rho=0$
in the above path integral. Varying $t$, corresponds to the finite gauge transformation
to the model we consider, where the calculations are much easier.}.

\subsection{Singular Polyakov Strings}

This picture corresponds to the boundary conditions 
fixing $b$. Then, the action is
\be
S_b=\int \left[\frac{1}{2\pi}p\dot x - a\dot b -\frac{\lambda}{2\pi} ba'
- \Lambda aa'\right]
\l{ba}\ee
We can fix $\lambda=2\pi$ just by rescaling $\tau\to \frac{2\pi}{\lambda}\tau$
and $\Lambda\to \frac{\lambda}{2\pi}\Lambda$.
Then, integrating over $a$ we get for the reduced system the action
\be
S_{SPS}=\frac{1}{2}\int (-\Lambda')^{-1}(\dot b+b')^2
\l{SPS}\ee
which is the Polyakov action for the string oscillations
$\frac{1}{2}\int\sqrt{g}g^{\alpha\beta}\partial_\alpha b\partial_\beta b$
with the singular world-sheet metric
$\sqrt{g}g^{\alpha\beta}=(-\Lambda')^{-1}\delta^\alpha_+\delta^\beta_+$.

The relevance of singular metrics in the context of gauge/string dualities was already 
considered in \cite{PolyakovS}.
Notice that the boundary gauge algebra is very simple and reads
\be
\left\{\oint ba', aa'\right\}_{P.B.}=(aa')'
\l{gaugealgebra}\ee
and the other brackets vanishing.
This can be easily shown to correspond to a definite contraction of the Virasoro
algebra (inequivalent to the tensionless limit one).
Expanding in Fourier modes $aa'(\sigma)=\sum_{n\not=0}l_ne^{in\sigma}$
and defining $l_0=\oint ba'$ (recall that this was already in the center of {\it another}
Virasoro before the tensionless contraction and generates displacements in $\sigma$),
we get $[l_0,l_n]=-nl_n$ and the others commuting.
This algebra follows from the usual Virasoro $[L_n,L_m]=(n-m)L_{n+m}$
by redefining $L_m=\kappa l_m$ for $m\not=0$ and $L_0=l_0$ and then letting $\kappa\to\infty$.
This contraction should correspond to the singular limit of the world-sheet metric.
Let us notice that the gauge algebra commutation relations are realized in terms 
of the differential operator $\partial_\sigma$ and the operator of multiplication by an arbitrary
periodic function in $[0,2\pi]$ without zero mode.
This acts on the space of the parameterization of the circle
and somehow should be related to the zig-zag symmetry transformations.

\subsection{Condensed strings}

This second picture is more convenient in order to compare with abelian gauge
theories and corresponds in keeping the string boundary conditions at fixed $a$.
Therefore the action is
\be
S_{a}=\int \left[\frac{1}{2\pi}p\dot x + b\dot a -\frac{\lambda}{2\pi} ba'
- \Lambda aa'\right]
\l{aa}\ee
As a first step we get rid of $\Lambda$ with a gauge transformation.
The action is in fact invariant under the gauge symmetry
\be
\delta_K a=0 \quad
\delta_K b=K'a-\frac{1}{2\pi}\oint K'a \quad
\delta_K \Lambda=\dot K-\frac{\lambda}{2\pi}K' \quad
\delta_K \lambda=0
\l{gt}\ee
which we can use to fix $\Lambda=0$. The second term in $\delta_K b$
is forced by the projection in the space of functions without zeromodes.
Notice that with the same procedure we can not fix also the 
constant Lagrange multiplier $\lambda$ since the gauge generator $\oint ba'$
corresponds to the following invariance of $S_{a}$
\be
\delta_k a= ka' \quad
\delta_k b= kb' \quad
\delta_k \Lambda=k \Lambda' \quad
\delta_k \lambda=0
\l{gt0}\ee
which are just constant shifts in $\sigma$.
As in the previous case, we rescale the hamiltonian time $\tau\to\frac{\lambda}{2\pi}\tau$
(and $\Lambda\to\frac{2\pi}{\lambda}\Lambda$ if we want to do it 
equivalently before fixing $\Lambda=0$) and now the action reads
\be
S_{a}=\int \left[\frac{1}{2\pi}p\dot x + b\dot a - ba'\right].
\l{aared}\ee
Now, integrating over $b$, we get the following familiar constraint on the string
oscillations
\be
\dot a - a'=0
\l{chiral}\ee
which fixes them to be chiral.
Eq.(\ref{chiral}) means that the string path-integral is restricted
to worldsheet such that
the time evolution of the string profile
is equivalent to a reparameterization of the string profile itself.
Therefore the evolution kernel vanishes for all the boundary conditions
incompatible with such a constrain, that is for all
but the ones for which the initial and the final string configurations
coincide up to a constant shift in $\sigma$, i.e. up to a gauge transformation
-- we have not fixed $\lambda$.
This means that in this string theory there is no dynamical evolution 
at all for the string configurations on the boundary.

\subsection{A gauge dual interpretation}

The picture we obtained in the last subsection has to be compared with $d-1$ dimensional Yang-Mills
gauge theory. By comparing with the results of 
\cite{massimo}, \cite{sw}, \cite{KP}, \cite{adscft}, \cite{also} and \cite{more}
we expect the gauge dual interpretation to be given in terms of an abelian
gauge theory seen as a zero coupling limit of a non-abelian one.

Our reasoning concerning the dual interpretation goes as follows.
We start by a strong version of string/gauge theory duality by which we 
mean that we expect that there exists a well defined string dual picture for 
YM theory in an AdS-like (i.e. non conformal in general) background,
the conformal cases being realized as conformal isometric (AdS$\times$ 
something, where ``something'' realizes extra global symmetries) backgrounds.
This duality is understood to be realized by a string theoretic
expression for the Wilson loops of the gauge theory as string worldsheet path
integrals with fixed boundary conditions on the loops locations.

In particular, therefore, we ask if also closed free tensionless
strings in AdS have a gauge theory dual. 

Since the theory is conformal invariant by construction (the BRST charge
commutes with all the generators of the conformal group), we expect the gauge
dual to be conformal too. 
Since the string theory is non supersymmetric, we believe 
that the gauge dual is neither and that, since the string theory is the
simplest conceivable, free and does not contain any length-scale,
the gauge dual has to be at a free point.
This already selects out abelian gauge theories, seen as zero coupling limits
of non-abelian ones, as unique candidates
\footnote{
We assume that pure non-abelian YM theory has no fixed point at intermediate
couplings. This situation 
in principle could be produced by adding some matter not necessarly in a
supersymmetric way. In this case, anyway, one expects some 
further global symmetry which would have no counterpart in the 
free closed tensionless string.}.

Therefore, our aim is to find if the boundary-boundary correlator 
for closed free tensionless strings has any gauge dual interpretation
as loop-operators expectation in an abelian gauge theory.
By the calculations done in the paper, the object at hand is just
the $\delta$-functional in the loop space between the initial and the final 
string boundary.

We identify the string (degenerate) propagator as the correlator between two charge conjugate
Wilson loops in the compact version at the infinite coupling point.
Showing this is simple
once the model of monopole condensation by Polyakov is
taken in consideration. I will shortly review it here for convenience
(see \cite{Polyakov97}).
The effective action describing the monopole resummation is given by
\begin{equation}
S=\int\left[\frac{1}{4e^2} B^2 +i\phi dB +e^2m^2 V(\phi)\right] +i\int_\Sigma B
\label{pa}\end{equation}
where $V(\phi)=1-\cos\phi$ and $m^2=m^2_0 e^{-e_0^2/e^2}$.
We take $\Sigma$ being a Riemann surface interpolating between two closed paths. 
In the path integral the sum over the surfaces is done as well as the
integration over $B$ and $\phi$.

It is clear that the perturbative expansion at small coupling
($e<<e_0 \Rightarrow e^2m^2\to0$) reproduces the standard perturbative abelian Wilson loop 
just by integration over $\phi$ which is enforcing the Bianchi identity on the
$2$-form $B$. This means that $B=dA$ and the sum over the surfaces factorises
since the coupling is effectivly only to the fixed boundary, that is
$\int_\Sigma B=\int_{\partial\Sigma}A$.

At intermediate couplings, the exact path integration reproduces the total
monopole sum of the compact abelian theory on top of the perturbative term.

At infinite coupling the potential term dominates and frezes $\phi=2n\pi$, where
$n$ is an arbitrary integer. This means that the integration on $\phi$ gets reduced to a 
discrete sum imposing then a quantization condition on $\int dB$.
The integration over the fluctuations of the $B$-field 
can be done by decomposing $B=dA+{}^*d^*V$, where $A$ is a 1-form and $V$ is a 3-form.
Notice that the quantization condition on $\int dB$ only involves $V$ and therefore
the integration along $A$ can be performed freely. This implies that
the $(d-3)$-form $\omega_\Sigma$ dual to the surface $\Sigma$ 
(that is such that $\int_\Sigma B=\int B\wedge\omega_\Sigma$) is closed. 
This condition can be satisfied only if the two boundary
components of $\Sigma$ coincide. So the surface sum reduces to closed surfaces such
that $d\omega_{\Sigma}=0$ and the path integral keeps $V$ only and all closed surfaces
to give an overall factor.

Therefore, we see that the tensionless free string boundary-boundary correlator exactly
coincides with the correlator of two conjugated Wilson loops in the compact
abelian gauge theory in the limit of infinite coupling.

Notice that the above Polyakov model strictly speaking has been worked out in
3 dimensions and generalized to 4.
The dual interpretation of the strongly coupled theory as a weakly coupled 
abelian $(d-4)$-form theory in any dimension $d-1$
is claimed to hold in \cite{Polyakov97} and in \cite{PolyakovS}.
Let us read it again for completeness starting again from the action (\ref{pa})
and lets keep the perturbative term $\frac{1}{4e^2}B^2$ while still freezing
the dynamic of $\phi$. 
The effective theory around the strong coupling point now is just
\begin{equation}
S=\int\left[\frac{1}{4e^2} B^2 +iB\wedge\omega_\Sigma\right]
\label{scpa}\end{equation}
and has to be compared with the first order formalism of an abelian $(d-4)$-form theory
with the small gauge coupling $\tilde e^2=\frac{1}{4e^2}$. 
As we concluded above, the summation over the surfaces is limited in this
regime to closed ones, that is such that $d\omega_\Sigma=0$.
This means that locally $\omega_\Sigma=dF$, where $F$ is a $(d-4)$-form.
Substituting in the action (\ref{scpa}) we get in fact
\begin{equation}
S=\int\left[\tilde e^2 B^2 +iB\wedge dF\right]
\label{fof}\end{equation}
and the summation over the closed surfaces gets replaced by the integral over the
local potential $F$. The action (\ref{fof}) is exactly the first order
formalism action of an abelian $(d-4)$-form theory in $d-1$ dimensions
with the (small) gauge coupling $\tilde e$. 

Notice that the last point, that is the equivalence of (\ref{scpa}) and
(\ref{fof}), has been proposed by Polyakov as a conjecture and
we do not claim we prove it here. We notice that if that Polyakov picture
is true, then it gives a natural gauge dual picture to the
tensionless string results.

Notice that for $d=5$, we find the usual (compact) Maxwell theory on $M_4$.
For $d=4$ we find free scalar fields in three dimensions. These were shown 
in \cite{KP} to be dual to a minimal higher spin theory in $AdS_4$. 

\section{Open Questions}

Let us first comment about conformal invariance.
This got apparently 
broken by our boundary conditions to Poincare' which is generated by
the $SO(2,d-1)$ generators fixing the null-direction we have chosen to 
implement the boundary reduction. 
Actually it has not been broken and its action can be calculated after the partial gauge 
fixing has been performed. The point is that it is just ``non manifest'' 
in the boundary picture.

Once this narrow corner (free and tensionless) is understood, one might pose the questions about 
how to resort the tension and how to switch on the string interactions.
It is clear that our technique seems specific of the tensionless limit of strings in AdS
and its generalization to the tensile case is not given for granted to be doable.
Moreover, the hamiltonian formalism is difficult to extend to the interacting 
first quantized string case
in general and it might be that one has to formulate the interacting theory 
in the second quantization, i.e. for string fields.

Linking our results with higher spin gauge theory is not too difficult
and naturally passes by the second quantization of the model.
Using the BRST charge for closed tensionless free strings in $AdS_d$
built in \cite{adscov}, one can consider the 
string field action $S=\frac{1}{2}<\Psi|\Omega\Psi>$ who's higher spin symmetry can be 
spectrally analyzed by means of its global symmetries.
Notice that the BRST charge for the closed tensionless strings, unlike the open string case, 
has the right (odd) ghost modes amount in order to the above action to be well defined.
The partial gauge fixing procedure that we performed here can be understood as
the fixation of a huge set of auxiliary fields along the path of \cite{MG}.
It would be very interesting to see how the theory looks like 
in detail after that procedure has been carried out.
This is based on the BRST charge implementing the remnant gauge symmetry. 
The higher spin symmetries of the resulting boundary reduced theory should be compared with 
the analogous ones in the dual abelian gauge theory
to further check the string/gauge correspondence for free tensionless strings. 
The second quantized tensionless theory seems to be the natural 
starting point for the introduction of string interactions.

Let us conclude by noticing that we realized the boundary reduction as a 
(partial) gauge fixing procedure. It would be interesting to understand if this 
mechanism is a peculiarity of the massless free case or if it could
be a way realize the holographic mechanism in more general cases.

\newpage

\ack

I would like to thank 
R.~Argurio, G.~Barnich, L.~Bonora and M.~Henneaux
for stimulating discussions.
I would like to thank also SISSA/ISAS, where this work was completed, for the kind hospitality. 
\noindent
This work is supported by the Marie Curie fellowship contract
HPMF-CT-2002-0185.
Work supported in part by the ``Actions de Recherche
Concert{\'e}es" of the ``Direction de la Recherche Scientifique -
Communaut{\'e} Fran{\c c}aise de Belgique", by 
the Belgian ``Interuniversity Attraction Poles
Programme -- Belgian Science Policy'', by IISN-Belgium
(convention 4.4505.86)  and by the European Commission RTN programme
HPRN-CT-00131, in which G.B. is associated to K. U. Leuven.

\end{document}